\documentclass[amssymb,amsmath,pra,floatfix,twocolumn]{revtex4}
\usepackage{graphicx}
\usepackage{color}
\usepackage{graphicx}
\usepackage{amssymb}
\usepackage{amsmath}
\usepackage{xypic}
\usepackage{array}

\begin{document}
\title{Programming quantum computers using 3-D puzzles, coffee cups, and doughnuts.}

\author{Simon J. Devitt}
\affiliation{Center for Emergent Matter Science, RIKEN, Wakoshi, Saitama 315-0198, Japan.}
\date{\today}

\begin{abstract}
The task of programming a quantum computer is just as strange as quantum mechanics itself.  But it now looks like a simple 3D puzzle may be the future tool of quantum software engineers.
\\
\\
{\em This is an article that appeared in the Fall 2016 issue of ACM-XRDS.  The reference list in this arXiv version has been expanded.}
\end{abstract}
\maketitle

\section{Introduction}

Andrew Steane, one of the pioneers of quantum computing, once reportedly quipped, 
\\
\\
{\bf {\em A quantum computer is an error correction machine - computation is just a byproduct.} }
\\
\\
Steane provides an extremely apt description of how any large-scale active quantum technology will ultimately behave.  Quantum information processing suffers from two disadvantages; controllable quantum bits, or qubits, are extremely susceptible to noise from bad control or the external environment, and quantum algorithms are, by nature, exceedingly sensitive to errors. Even a single error during the execution of an algorithm can lead to essentially random output.  

Hence, Quantum Error Correction (QEC) was quickly recognized as a necessity for any commercially viable computational or communications protocol, and the theoretical development of error correction techniques is as old as the first architectural models for quantum computers \cite{CZ95,K98,LD98,KLM01}.  Developed in the mid 1990's by researchers such as Peter Shor, Andrew Steane, Alexei Kitaev, Daniel Gottesman and Robert Calderbank \cite{CLSZ95,S95,BDSW96,S96,CRSS98}, QEC, when combined with the principle of fault-tolerant quantum computation \cite{DS96,S96+,K97,G98,G00+,K03,G09,DMN13}, leads to arguably the most important theoretical result in quantum computing: the threshold theorem \cite{AB97}.
\\
\\
{\bf {\em A quantum computation of arbitrary size can be completed successfully with faulty qubits, with a polylogarithmic resource overhead, provided that the physical error rate associated with each qubit and logic gate is below a maximum value, dubbed the fault-tolerant threshold. } }
\\
\\
What this theorem is basically saying is that provided the error experienced by each qubit is below a certain value (the threshold), error correction will correct more errors than it introduces and a computation, no matter how large, will always be error free by introducing extra qubits.

The value of the fault-tolerant threshold is determined by many factors - the QEC code utilized, the way error correction codes are constructed, and any physical restrictions of the quantum hardware such as if qubits can be coupled together arbitrarily or are interactions restricted to a fixed geometry \cite{KLZ96,AGP08,KBM09,AT06,AP08,PR11,SDT07,DA07,SFH07,S03,SBFRYSGF06}.  Initial estimates were very unfavorable, with thresholds for the Steane code and others of the order of 0.01\% or lower  \cite{KLZ96,K03,AB97,K97,G97+}. However, this has improved dramatically with the development of topological models of QE. These exhibit fault-tolerant thresholds approaching 1\% for models such as the surface code \cite{DKLP02,FSG08,FMMC12,S14,WFSH09}, and Raussendorf code \cite{RHG06,RH07,RHG07}. These codes are also much more amenable to physical implementation as they are defined on a two-dimensional (surface code) or three-dimensional (Raussendorf code) array of nearest-neighbor interacting qubits.  

The high fault-tolerant threshold, the nearest-neighbour nature of these topological codes, and the way in which quantum algorithms are implemented have resulted in them becoming the preferred technique for large-scale quantum computing architectures \cite{DFSG08,DMN08,D09,JMFMKLY10,YJG10,NTDS13,MRRBMD14,LBS10,LHMB15,LWFMDWH15,HPHHFRSH15,ONRMB16}.  Essentially all major physical system are now targeting either the surface code or Raussendorf code for their architectures, and physical systems such as Ion Traps and Superconducting qubits are now demonstrating gate and qubit error rates either at, or below, the fault-tolerant threshold \cite{B14,L16}.  It is becoming increasingly probable that a functional, commercial quantum computing system will be build using topological QEC as the fundamental computational model.  

In Ref.\cite{PFW16}, Paler, Fowler and Wille have compiled a review that details how both computation and error correction is performed in these models.  This article continues from this review and will examine both the structure of a topological quantum circuit and how these circuits will ultimately be optimised and implemented on a real world quantum computer. 

\section{Topology: Cofffee cups and doughnuts}

Topology, unsurprisingly, plays a crucial role in the function and operation of topological quantum error correction.  As with essentially all QEC codes that are considered implementable on large-scale hardware, topological quantum codes are defined in terms of stabiliser operators \cite{G98+}.  A quantum state $|\psi\rangle$ is stabilized by an operator $K$, such that $K|\psi\rangle = |\psi\rangle$.  A topological quantum code is defined by a set of these operators, which are defined locally. That is, they are defined over a small group of qubits that are nearby to each other\cite{K03}.  However, the encoded state defined by these operators have certain global properties.   Logical operations, those that define the encoded qubit state, are defined with respect to the entire state - they cannot be defined locally.  This is the essential nature of a topological code.  Individual stabilizers that are used to perform error correction are defined locally, while logical information is defined globally.  

As summarized in Ref. \cite{PFW16}, a two-dimensional lattice of qubits (for the surface code), or a three-dimensional lattice of qubits (for the Raussendorf code) defines a unique quantum state \cite{FMMC12,RHG07}.  The eigenvalues of each of the stabilizers associated with the lattices are measured in order to detect and correct for quantum errors that can occur due to imperfect physical qubits and gates \cite{DFTMN10,GP10,GP14,F15+}.  Information is encoded into this lattice through the creation of holes, or defects.  Defects are regions of the lattice that have been deactivated by having qubits in these regions removed.  By removing qubits, or deactivating certain parts of the lattice, degrees of freedom are introduced into the quantum state that can be used to store and manipulate information which is protected from errors due to the properties of the remaining lattice, which is called the bulk.  

Interactions (quantum gates) in this model are enacted through an operation called braiding.  Braiding is where defects are perturbed such that they {\em move} through the lattice as it evolves in time, and wrap around each other like a tangled ball of string \cite{FMMC12,PF13,DN12}.  An example of a large topological circuit that enacts a set of logical gates on encoded qubits, in a fault-tolerant way, is illustrated in Figure \ref{fig:four}.  The spatial cross section is illustrated, as well as the temporal axis.  The spatial cross section defines the number of qubits used in the surface code while the temporal axis defines how defects are created and manipulated over time \cite{DSMN13}.  In the Raussendorf model, all three dimensions of the lattice consist of physical qubits which are sequentially measured along the temporal axis.  Measurements are used to define and manipulate the defects through teleportation along this temporal axis of the Raussendorf code. 

\begin{figure*}[ht!]
\begin{center}
\resizebox{0.8\linewidth}{!}{0.4\includegraphics{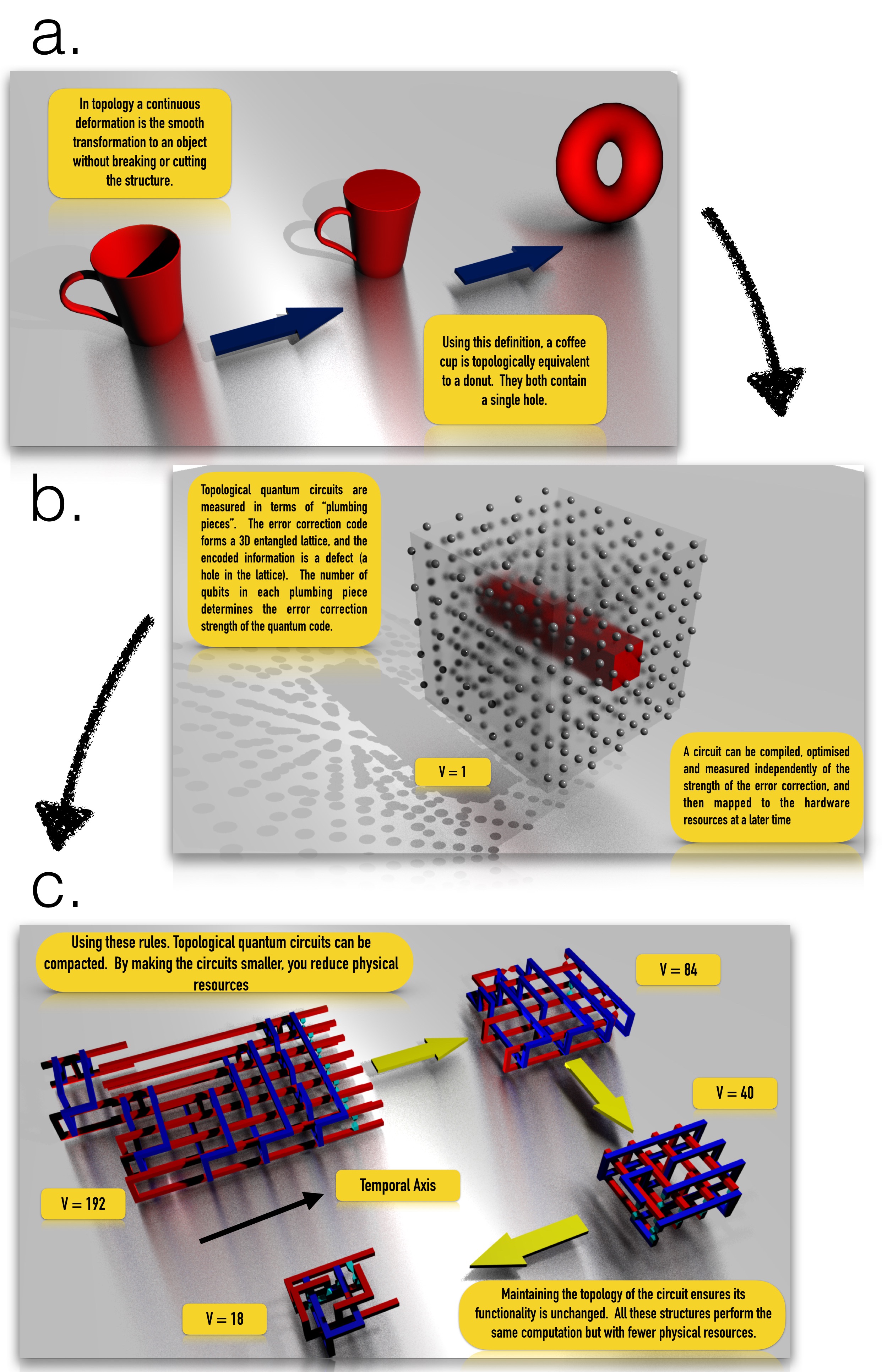}}
\end{center}
\caption{{\bf How to optimise topological quantum circuits.}  Figure a. represents the topological equivalency of a coffee cup and a donut.  Figure b. illustrates a {\em plumbing piece}, the basic unit of measure for a topological quantum circuit. Increasing the error correction code required more physical qubits for each pluming piece.  Figure c.  Topological deformation can be used to reduce the volume, and hence physical resources, without changing its information processing properties.}
\label{fig:one}
\end{figure*}

These structures are topological in nature and, hence, standard definitions of topology apply.  The nature of a topological space is that it is preserved through operations known as continuous deformation.  Continuous deformation is where a structure is stretched or bent without being cut or glued together at any point.  The quintessential example of this is the topological equivalence between a coffee cup and a donut shown in Figure \ref{fig:one}a.  Simply by stretching and bending, the coffee cup can be converted to a donut and vice versa.  As each structure has only a single hole, they are topologically equivalent.  

A qualtum program is therefore, literally, defined and described by a puzzle and this puzzle can be shaped, stretched and molded to change the physical resources needed by a quantum program without changing the program itself \cite{PF13}.

\section{Measuring and benchmarking quantum circuits}

In order to derive relevant metrics when constructing, compiling, and optimizing topologically error-corrected circuits we need to understand how a circuit relates to the number of qubits and the physical computational time when they are implemented \cite{DSMN13}.  Regardless of whether we are talking about the Surface code or the Raussendorf lattice, the relationship between a topological circuit and physical resources is identical.  The fundamental unit of measure is illustrated in Figure \ref{fig:one}b.  A {\em plumbing piece} of a topological quantum circuit is a three-dimensional cubical volume that has an edge length related to the desired strength of the underlying quantum code.  For a distance d code (sufficient to correct upto $t = (d-1)/2$ errors), this plumbing piece has an edge length containing $5d/4$ plaquette cells for the surface code),or $5d/4$ cells in the Raussendorf lattice.  At the centre of this plumbing piece is the defect, which has a circumference of $d$ plaquettes.  Figure \ref{fig:one}b illustrates an example for $d = 4$.  The plumbing piece gives a scale independent factor to allow us to measure topological quantum circuits without having to specify the strength of the underlying error correction \cite{FD12,FDJ13}.  

Using topological circuit volumes in terms of plumbing pieces allows us to directly calculate the total number of qubits and computational time.  For the surface code, the plumbing piece requires a total of $Q = 25d^2/4 + 5d + 1$ qubits and $T = 5d/4$ steps.  Each step is defined as a syndrome extraction circuit for both bit-errors and phase-errors. For the Raussendorf lattice, a plumbing piece requires a total of $Q = 6d^3+9d^2+3d$ qubits. For a larger topological circuit, we can use their volume to first calculate the required strength of error correction d to ensure no logical errors occur during implementation, and then calculate the total number of resources needed by converting the volume to physical qubit numbers and computational time.  

This method of designing quantum programs is very useful as we do not need to redesign anything about the actual quantum hardware when we change the quantum program.  We just need to make sure we have enough qubits to do the job.

\section{Constructing and compiling initial topological circuits}

\begin{figure*}[ht!]
\begin{center}
\resizebox{0.8\linewidth}{!}{0.4\includegraphics{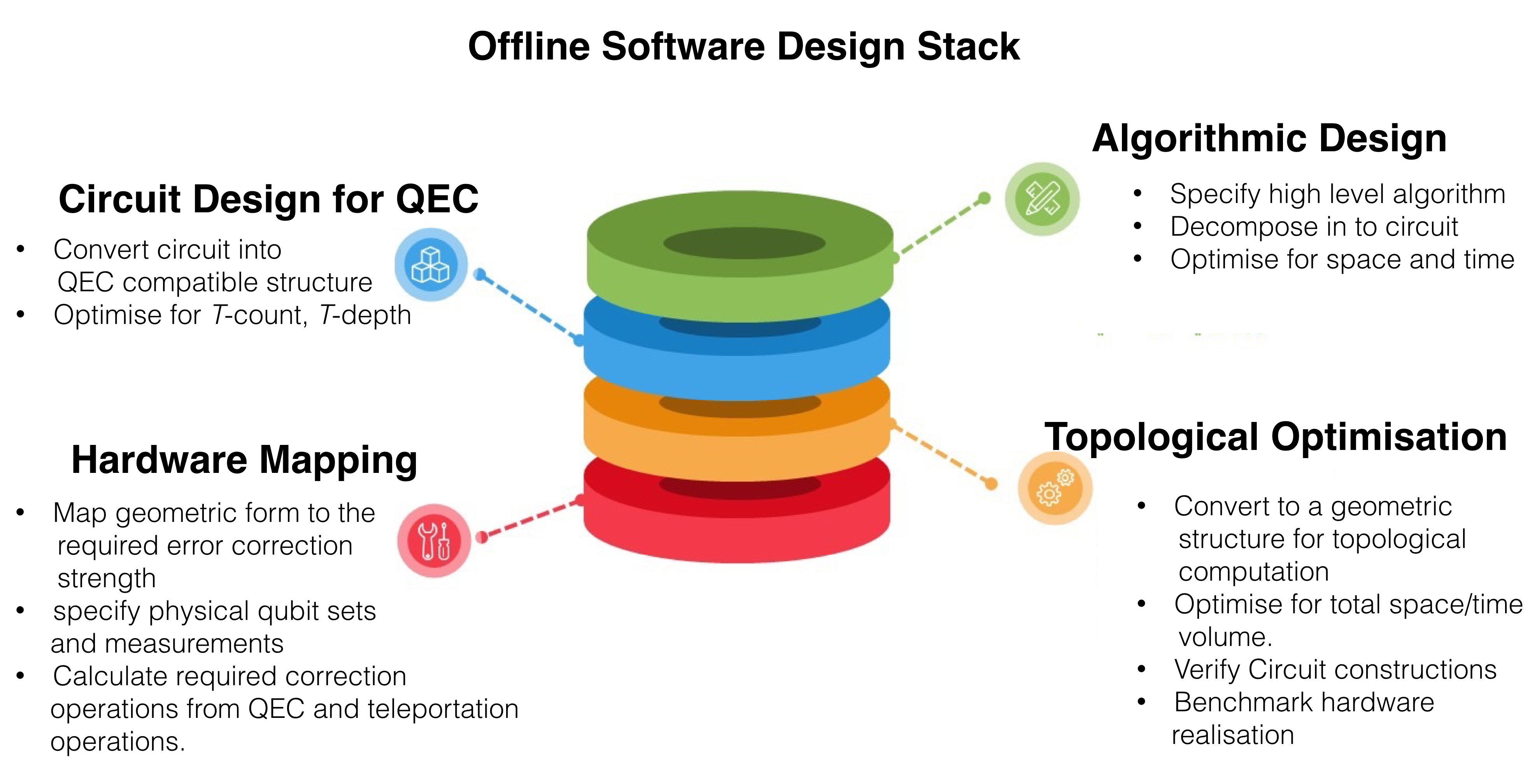}}
\end{center}
\caption{{\bf Offline compilation and optimisation stack \cite{MD16}}.  The conversion of a abstract high level quantum algorithm to an unoptimised topological form. What is addressed in this review is the third level which corresponds to topologically compacting a given circuit specification prior to hardware implementation.}
\label{fig:two}
\end{figure*}

Before a given computation can be suitably optimized in the topological formalism, quantum circuits need to be compiled and constructed from the original algorithmic specification \cite{G06,GLRSV13,WS14}.  Figure \ref{fig:two} illustrates the broad structure of the compilation stack needed for an arbitrary algorithm.  The stack is partitioned into several stages;

Algorithm to Circuit: A quantum circuit, consisting of single, two and three qubit gate primitives, is derived from the abstract algorithm.  This circuit can be optimized for depth, and/or number of qubits \cite{Z98,VI05,WC00,AMMR12,WBCB14}. 
Circuit to fault-tolerant primitives:  The abstract circuit is further decomposed into gate sets that have well defined fault-tolerant implementations in the topological code.  Again, optimizations can occur at this level \cite{GS12,J13+,J13+++,RS14,GKMR14,KMM13}
Fault-tolerant circuit to ICM form:  The circuit consisting of fault-tolerant gate primitives is then converted to a form called Initialization, Controlled-Not, Measurement (ICM).  This allows us to build in appropriate auxiliary protocols needed before explicitly converting them to a topological implementation \cite{PPND15}.
Canonical Topological Form:  Once written in ICM form, the circuit can be converted to an unoptimized canonical form in the topological model, prior to further resource optimization \cite{PDF16}.

The conversion of a higher level circuit to a canonical topological form is a complicated but well defined process, and researchers have designed several software packages to perform this task.

\begin{figure*}[ht!]
\begin{center}
\resizebox{0.9\linewidth}{!}{0.4\includegraphics{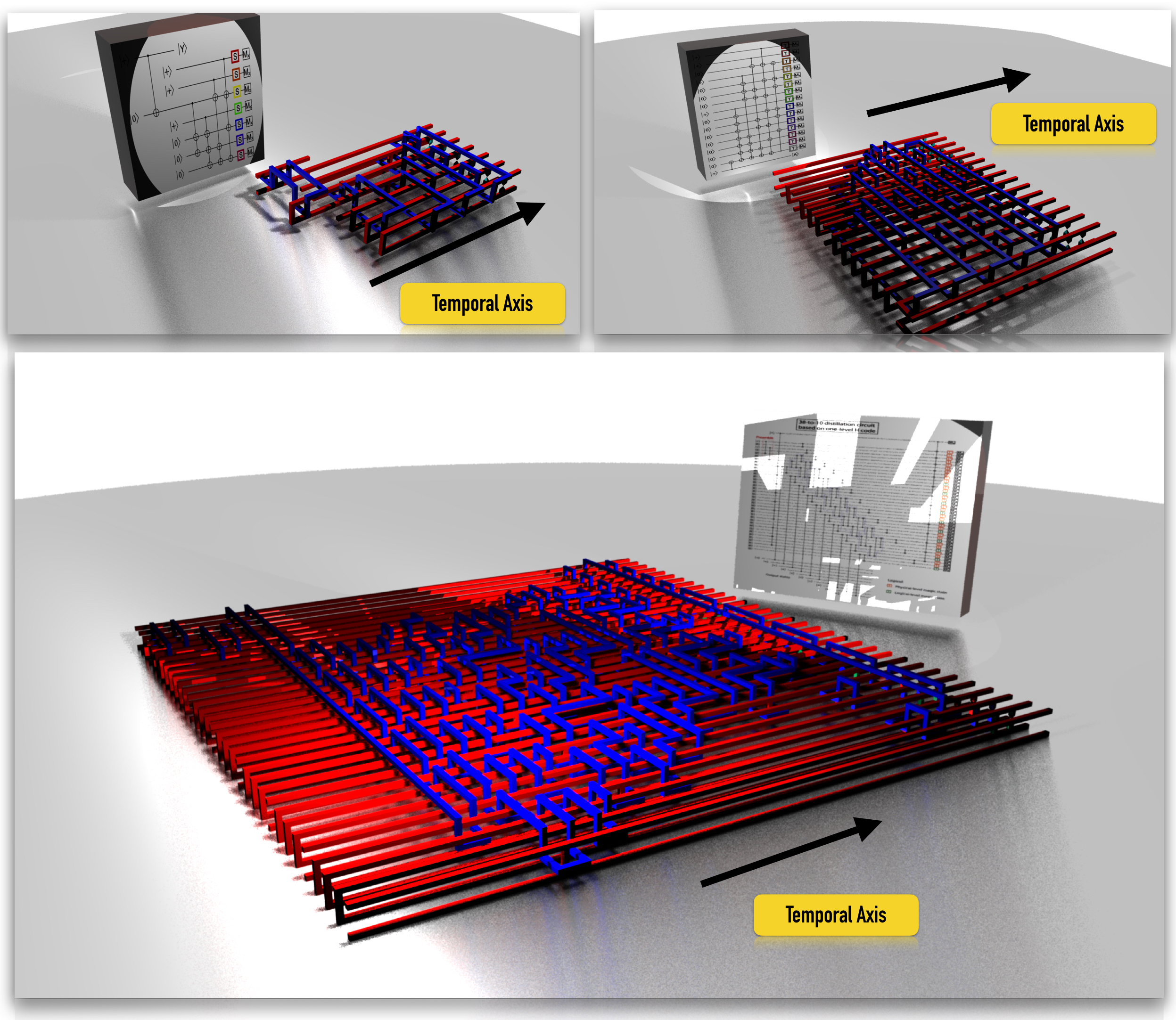}}
\end{center}
\caption{{\bf Canonical topological quantum circuits.}  In each figure we illustrate a quantum circuit (written in the standard pictorial form) and the corresponding, unoptimised topological quantum circuit.  Each circuit can be measured in terms of volume. The temporal axis is defined as the temporal evolution of this structure as the computation proceeds}
\label{fig:four}
\end{figure*}

In Figure \ref{fig:four} we illustrate several examples of a canonical topological form and the quantum circuits they are derived from \cite{BH12,J13++}.  Each of the circuits shown are known as magic state distillation circuits \cite{BK05+} and are used to enact certain gates that require high-fidelity ancillary states.  Each of these circuits have a corresponding volume and, hence, can be used to estimate physical resources.  

\section{Topological Optimisation}

The next crucial step in the design stack for error-corrected quantum circuits is topological optimization \cite{PF13}. This also happens to be the most underdeveloped area of the stack.  Almost all other elements have been completely understood or are being heavily researched, including efficient methods for circuit construction and optimization, and fault-tolerance.  Although it has received little attention, there are strong reasons to believe that topological optimization may result in some of the largest resource savings if implemented well \cite{FD12,FDJ13}.  

The basic principal is illustrated in Figure \ref{fig:one}c.  The canonical circuit begins with a volume of $V=192$ plumbing pieces, and in the same way as a coffee cup can be deformed topologically into a donut, we can slowly compact the physical three dimensional volume of the circuit without altering its computational function.  There are some additional rules not related to continuous deformation, and unique to the computational model, one of which is called bridging, that can be used to reduce the physical volume of a topological circuit significantly \cite{FD12}.  After many steps (not illustrated), the final volume of the topological circuit is reduced to $V=18$ - over an order of magnitude smaller than the original canonical form.  This amount of optimization is indeed significant.  For a surface code quantum computer, the number of qubits required for implementation can be reduced by orders of magnitude by simply compressing these structures.  However, two theoretical questions still remain unanswered.  

The first is to provide a lower bound, or exact definition for optimality, for a topological circuit.  While we can compress, we do not have a condition for optimality given the original circuit specification.  The second question concerns the classical complexity of the algorithm required to find this optimal solution.  While this problem does appear closely related to the three dimensional bin packing problem \cite{Bin00}, which is known to fall into the complexity class of NP-Hard, there are small differences in the topological QEC model that may imply that these two problems do not directly map on to each other. It is still possible that the optimization of topological quantum circuits may be provably classically efficient to calculate.  

Circuits that have been currency compacted have been done so manually. This is obviously not a viable approach for large-scale implementations of error-corrected quantum algorithms.  There have been very small steps to try and build automated topological optimization packages \cite{PF13}, but these have, so far, only illustrated the potential difficulties in creating the required software.  Being able to optimize even moderately large quantum circuits will not be possible without automated software, and it appears as though techniques in machine learning and artificial Intelligence may be required to provide resource efficient solutions. 

\section{meQuanics: The Quantum Computer Game} 

The approach that we recently took to address this problem was inspired by projects in the biological sciences \cite{foldit,eyewire} that attempt to solve scientifically useful but difficult problems by utilizing the computing capacity of the general public.  This technique, sometimes referred to as Citizen Science, was pioneered by projects such as FoldIt (which aimed to find the three-dimensional structure of biological proteins given their constituent sequence of amino acids), and Eyewire (designed to map neural connections in the retina) and has achieved significant success. 

Given the relatively simple 3D puzzle structure of topological quantum circuits, and the simple success metric of minimal physical volume, we have tried the same approach.  An initial prototype of a platform we have dubbed meQuanics [www.mequanics.com.au], designed to convert the topological optimization problem into a simple 3D puzzle game, has been released online. Designed for touch-based platforms such as smartphones and tablet devices, meQuanics creates an online social media environment where the general public can compete and collaborate to find small volume solutions to various quantum sub-circuits that are critical for large-scale quantum computation.  

\begin{figure*}[ht!]
\begin{center}
\resizebox{0.85\linewidth}{!}{0.4\includegraphics{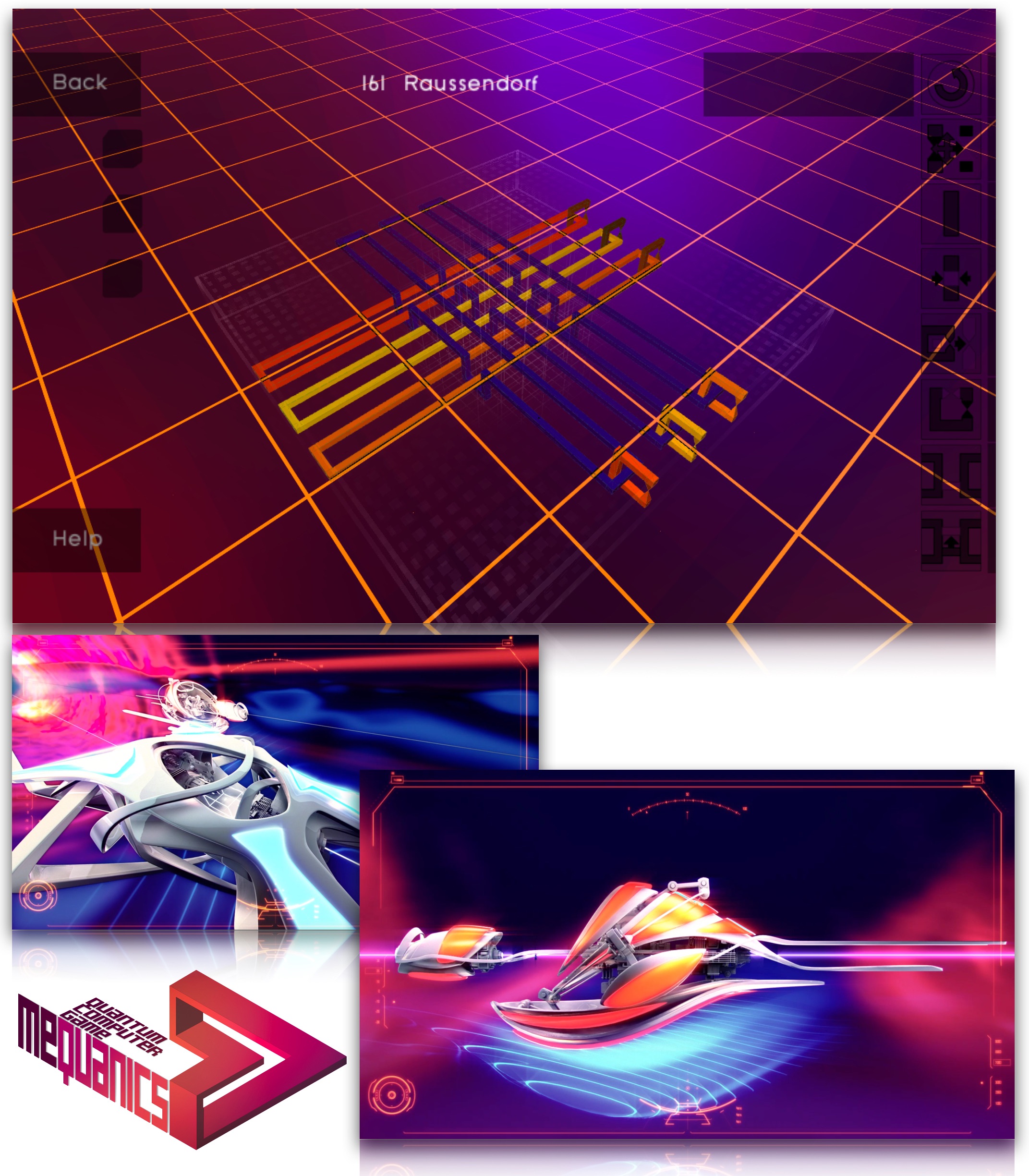}}
\end{center}
\caption{{\bf Current assets for the prototype client of meQuanics}. Here we show a current screenshot from the meQuanics client [www.mequanics.com.au] and some digital assets that will be used for the final game client.}
\label{fig:three}
\end{figure*}

While it is conceivable that users can derive compact solutions that are significantly smaller than solutions we currently have, the primary goal is not the solutions but rather the process individual players use to generate these solutions.  It is well known within the machine learning and AI community that the success of these techniques requires a database of information that the machine system can use to learn.  While for many problems there is an existing database of material that can be utilized (such as the AlphaGo platform of the DeepMind project at Google), for this particular problem there are, essentially, no previous examples that we can use to train an appropriate automated program.  

While the prototyping stage has demonstrated proof-of-principle client, there is still significant development required.  The most important goal is to create a game that is competitive in the larger marketplace of mobile and touch-based games. While the novelty of a game that is very closely related to quantum computing development may induce a large number of users to try it out, long term retention of gamers is required to generate the necessary data sets for the project to be successful.  

Gameplay and social interaction in meQuanics is now a major focus of development.  The basic narrative is that of an interstellar race, where each ship is powered by a quantum computer.  By minimizing the volume of puzzles, players increase their speed and ultimately win over other players working on the same problem.  The online interaction environment is being designed under the assumption that each client continuously updates information to central servers informing us how players are tackling problems.  Individual players can take a continuously expanding solution tree that begins from a specific canonical circuit structure and either improve on other players solutions, or backtrack and proceed down a different pathway that could lead to better solutions.  Elements of the social media and gameplay environment are illustrated in Figure \ref{fig:three}.

\section{The Future}
  
Future prospects on the software component of large-scale quantum technologies is promising.  Not only is there a vast amount of unsolved problems that can be addressed, even by researchers not heavily trained in quantum physics, but the theoretical similarities of essentially all major experimental hardware models implies that software solutions are applicable to all systems.  This is now evident from the founding of four private start-up companies, exclusively focused on the software component of quantum technology: QxBranch [www.qxbranch.com.au], 1Qbit [www.1Qbit.com] QCware [www.qcware.com] and Cambridge Quantum Computing [www.cambridgequantum.com]

While each element of the software compilation stack has been addressed at some level, a functional quantum computer will require a completely integrated set of classical software compilation and optimization packages.  The expertise of the classical software engineering community will be vital to this.  While physicists may be the experts in building quantum hardware, efficient and reliable software control will probably be developed by those already well versed in the advanced techniques of classical software engineering.  The fact that these problems are not intrinsically quantum in nature will make it easier for those without explicit training in quantum physics to get involved and make important contributions in this arena.

Quantum information technology is currently experiencing a second renaissance in advancement and investment from both the public and private sectors.  As such, there is consensus amongst experts that it is no longer a question of if a large-scale quantum computer can be built, but when.  The quantum revolution has the potential to be as significant as the digital revolution of the 20th century, and there is now a worldwide race to be the first to show a commercial advantage in deploying large-scale computers, communication networks, sensors, and other active quantum technology.  We stand at the cusp of an exciting new age in computing, with a significant laundry list of problems to interest pioneers. 

\section{Acknowledgements}

We would like to thank A. Paler and A.G. Fowler for collaborating on research summarized here.  meQuanics was developed in collaboration with K. Nemoto, K. Bruegmann, E. Gray, P. Daouadi, M. Everitt, Y. Quemener and F. Schittig through funds provided by the Hayao Nakayama Foundation for Science and Technology and Culture and the Japanese National Institute of Informatics. S.J. Devitt acknowledges support from the JSPS grant for challenging exploratory research and the JST ImPact project.

\clearpage

\bibliographystyle{unsrt}
\bibliography{XRDS.bib} 

\begin{thebibliography}{10}

\bibitem{CZ95}
J.I. Cirac and P.~Zoller.
\newblock {Quantum Computations with Cold Trapped Ions}.
\newblock {\em Phys. Rev. Lett.}, 74:4091, 1995.

\bibitem{K98}
B.E. Kane.
\newblock {A Silicon-Based nuclear spin Quantum Computer}.
\newblock {\em Nature (London)}, 393:133, 1998.

\bibitem{LD98}
D.~Loss and D.P. DiVincenzo.
\newblock {Quantum Computation with quantum dots}.
\newblock {\em Phys. Rev. A.}, 57:120, 1998.

\bibitem{KLM01}
E.~Knill, R.~Laflamme, and G.J. Milburn.
\newblock {A Scheme for Efficient Quantum Computation with linear optics}.
\newblock {\em Nature (London)}, 409:46, 2001.

\bibitem{CLSZ95}
I.L. Chuang, R.~Laflamme, P.W. Shor, and W.H. Zurek.
\newblock {Quantum Computers, factoring, and decoherence}.
\newblock {\em Science}, 270:1633, 1995.

\bibitem{S95}
P.W. Shor.
\newblock {Scheme for reducing decoherence in quantum computer memory}.
\newblock {\em Phys. Rev. A.}, 52:R2493, 1995.

\bibitem{BDSW96}
C.H. Bennett, D.P. DiVincenzo~J.A. Smolin, and W.K. Wooters.
\newblock {Mixed State Entanglement and Quantum Error Correction}.
\newblock {\em Phys. Rev. A.}, 54:3824, 1996.

\bibitem{S96}
A.M. Steane.
\newblock {Error Correcting Codes in Quantum Theory}.
\newblock {\em Phys. Rev. Lett.}, 77:793, 1996.

\bibitem{CRSS98}
A.R. Calderbank, E.M. Rains, P.W. Shor, and N.J.A. Sloane.
\newblock {Quantum Error Correction via Codes Over GF(4)}.
\newblock {\em IEEE Trans. Inform. Theory}, 44:1369, 1998.

\bibitem{DS96}
D.P. DiVincenzo and P.W. Shor.
\newblock {Fault-Tolerant error correction with efficient quantum codes}.
\newblock {\em Phys. Rev. Lett.}, 77:3260, 1996.

\bibitem{S96+}
P.W. Shor.
\newblock {Fault-Tolerant quantum computation}.
\newblock {\em Proc. 37th IEEE Symp. on Foundations of Computer Science.},
  pages 56--65, 1996.

\bibitem{K97}
A.Y. Kitaev.
\newblock {Quantum Computations: algorithms and error correction}.
\newblock {\em Russ. Math. Serv.}, 52:1191, 1997.

\bibitem{G98}
D.~Gottesman.
\newblock {A theory of Fault-Tolerant quantum computation}.
\newblock {\em Phys. Rev. A.}, 57:127, 1998.

\bibitem{G00+}
D.~Gottesman.
\newblock {Fault-Tolerant Quantum Computation with Local Gates}.
\newblock {\em J. Mod. Opt.}, 47:333, 2000.

\bibitem{K03}
A.~Kitaev.
\newblock {Fault-tolerant quantum computation by anyons}.
\newblock {\em Ann. Phys.}, 303:2, 2003.

\bibitem{G09}
D.~Gottesman.
\newblock {An Introduction to Quantum Error Correction and Fault-Tolerant
  Quantum Computation}.
\newblock {\em arXiv:0904.2557}, 2009.

\bibitem{DMN13}
S.J. Devitt, W.J. Munro, and K.~Nemoto.
\newblock Quantum error correction for beginners.
\newblock {\em Rep. Prog. Phys.}, 76:076001, 2013.

\bibitem{AB97}
D.~Aharonov and M.~Ben-Or.
\newblock {Fault-tolerant Quantum Computation with constant error}.
\newblock {\em Proceedings of 29th Annual ACM Symposium on Theory of
  Computing}, page~46, 1997.

\bibitem{KLZ96}
E.~Knill, R.~Laflamme, and W.H. Zurek.
\newblock {Accuracy threshold for Quantum Computation}.
\newblock {\em quant-ph/9610011}, 1996.

\bibitem{AGP08}
P.~Aliferis, D.~Gottesman, and J.~Preskill.
\newblock {Accuracy threshold for Quantum Computation}.
\newblock {\em Quant. Inf. Proc.}, 8:181--244, 2008.

\bibitem{KBM09}
H.G. Katzgraber, H.~Bombin, and M.A. Martin-Delgato.
\newblock {Error Threshold for Color Codes and Random 3-Body Ising Models}.
\newblock {\em Phys. Rev. Lett.}, 103:090501, 2009.

\bibitem{AT06}
P.~Aliferis and B.M. Terhal.
\newblock {Fault-tolerant quantum computation for local leakage faults}.
\newblock {\em Quant. Inf. Comp.}, 7:139, 2007.

\bibitem{AP08}
P.~Aliferis and J.~Preskill.
\newblock {Fault-Tolerant quantum computation against biased noise}.
\newblock {\em Phys. Rev. A.}, 78:052331, 2008.

\bibitem{PR11}
A.~Paetznick and B.W. Reichardt.
\newblock {Fault-tolerant ancilla preparation and noise threshold lower bound
  for the 23-qubit Golay code}.
\newblock {\em Quant. Inf. Comp.}, 12:1034, 2012.

\bibitem{SDT07}
K.M. Svore, D.P. DiVincenzo, and B.M. Terhal.
\newblock {Noise Threshold for a Fault-Tolerant Two-Dimensional Lattice
  Architecture}.
\newblock {\em Quant. Inf. Comp.}, 7:297, 2007.

\bibitem{DA07}
D.P. DiVincenzo and P.~Aliferis.
\newblock {Effective Fault-Tolerant Quantum Computation with slow measurement}.
\newblock {\em Phys. Rev. Lett.}, 98:020501, 2007.

\bibitem{SFH07}
A.~Stephens, A.G. Fowler, and L.C.L. Hollenberg.
\newblock {Universal Fault-Tolerant Computation on bilinear nearest neighbor
  arrays}.
\newblock {\em Quant. Inf. Comp.}, 8:330, 2008.

\bibitem{S03}
A.M. Steane.
\newblock {Overhead and noise threshold of Fault-Tolerant quantum error
  correction}.
\newblock {\em Phys. Rev. A.}, 68:042322, 2003.

\bibitem{SBFRYSGF06}
T.~Szkopek, P.O. Boykin, H.~Fan, V.P. Roychowdhury, E.~Yablonovitch, G.~Simms,
  M.~Gyure, and B.~Fong.
\newblock {Threshold Error Penalty for Fault-Tolerant Computation with Nearest
  Neighbour Communication}.
\newblock {\em IEEE Trans. Nano.}, 5:42, 2006.

\bibitem{G97+}
D.~Gottesman.
\newblock {PhD Thesis (Caltech)}.
\newblock {\em quant-ph/9705052}, 1997.

\bibitem{DKLP02}
E.~Dennis, A.~Kitaev, A.~Landahl, and J.~Preskill.
\newblock {Topological Quantum Memory}.
\newblock {\em J. Math. Phys.}, 43:4452, 2002.

\bibitem{FSG08}
A.G. Fowler, A.M. Stephens, and P.~Groszkowski.
\newblock {High threshold universal quantum computation on the surface code}.
\newblock {\em Phys. Rev. A.}, 80:052312, 2009.

\bibitem{FMMC12}
A.G. Fowler, M.~Mariantoni, J.M. Martinis, and A.N. Cleland.
\newblock {Surface codes: Towards practical large-scale quantum computation}.
\newblock {\em Phys. Rev. A.}, 86:032324, 2012.

\bibitem{S14}
A.M. Stephens.
\newblock {Fault-tolerant thresholds for quantum error correction with the
  surface code}.
\newblock {\em Phys. Rev. A.}, 89:022321, 2014.

\bibitem{WFSH09}
D.S. Wang, A.G. Fowler, A.M. Stephens, and L.C.L. Hollenberg.
\newblock {Threshold Error rates for the toric and surface codes}.
\newblock {\em Quant. Inf. Comp.}, 10:456, 2010.

\bibitem{RHG06}
R.~Raussendorf, J.~Harrington, and K.~Goyal.
\newblock {A Fault-tolerant one way quantum computer}.
\newblock {\em Ann. Phys.}, 321:2242, 2006.

\bibitem{RH07}
R.~Raussendorf and J.~Harrington.
\newblock {Fault-tolerant quantum computation with high threshold in two
  dimensions}.
\newblock {\em Phys. Rev. Lett.}, 98:190504, 2007.

\bibitem{RHG07}
R.~Raussendorf, J.~Harrington, and K.~Goyal.
\newblock {Topological fault-tolerance in cluster state quantum computation}.
\newblock {\em New J. Phys.}, 9:199, 2007.

\bibitem{DFSG08}
S.J. Devitt, A.G. Fowler, A.M. Stephens, A.D. Greentree, L.C.L. Hollenberg,
  W.J. Munro, and K.~Nemoto.
\newblock {Architectural design for a topological cluster state quantum
  computer}.
\newblock {\em New. J. Phys.}, 11:083032, 2009.

\bibitem{DMN08}
S.J. Devitt, W.J. Munro, and K.~Nemoto.
\newblock {High Performance Quantum Computing}.
\newblock {\em arXiv:0810.2444, Prog. Informatics}, 8:1--7, 2011.

\bibitem{D09}
D.P. DiVincenzo.
\newblock {Fault-tolerant architectures for superconducting qubits}.
\newblock {\em Phys. Scr.}, T137, 2009.

\bibitem{JMFMKLY10}
N.~Cody Jones, R.~Van Meter, A.G. Fowler, P.L. McMahon, J.~Kim, T.D. Ladd, and
  Y.~Yamamoto.
\newblock {A Layered Architecture for Quantum Computing Using Quantum Dots}.
\newblock {\em Phys. Rev. X.}, 2:031007, 2012.

\bibitem{YJG10}
N.Y. Yao, L.~Jiang, A.V. Gorshkov, P.C. Maurer, G.~Giedke, J.I. Cirac, and M.D.
  Lukin.
\newblock {Scalable Architecture for a Room Temperature Solid-State Quantum
  Information Processor}.
\newblock {\em Nature Communications}, 3:800, 2012.

\bibitem{NTDS13}
K.~Nemoto, M.~Trupke, S.J. Devitt, A.M. Stephens, K.~Buczak, T.~Nobauer, M.S.
  Everitt, J.~Schmiedmayer, and W.J. Munro.
\newblock {Photonic architecture for scalable quantum information processing in
  NV-diamond}.
\newblock {\em Phys. Rev. X.}, 4:031022, 2013.

\bibitem{MRRBMD14}
C.~Monroe, R.~Raussendorf, A.~Ruthven, K.R. Brown, P.~Maunz, L.-M. Duan, and
  J.~Kim.
\newblock {Large Scale Modular Quantum Computer Architecture with Atomic Memory
  and Photonic Interconnects}.
\newblock {\em Phys. Rev. A.}, 89:022317, 2014.

\bibitem{LBS10}
Ying Li, Sean~D. Barrett, Thomas~M. Stace, and Simon~C. Benjamin.
\newblock Fault tolerant quantum computation with nondeterministic gates.
\newblock {\em Physical Review Letters}, 105(25):250502--, 12 2010.

\bibitem{LHMB15}
Y.~Li, P.C. Humphreys, G.J. Mendoza, and S.C. Benjamin.
\newblock {Resource Costs for Fault-Tolerant Linear Optical Quantum Computing}.
\newblock {\em Phys. Rev. X.}, 5:041007, 2015.

\bibitem{LWFMDWH15}
B.~Lekitsch, S.~Weidt, A.G. Fowler, K.~M{\o o}lmer, S.J. Devitt, C.~Wunderlich,
  and W.K. Hensinger.
\newblock {Blueprint for a microwave ion trap quantum computer}.
\newblock {\em arxiv:1508.00420}, 2015.

\bibitem{HPHHFRSH15}
C.D. Hill, E.~Peretz, S.J. Hile, M.G. House, M.~Fuechsle, S.~Rogge, M.Y.
  Simmons, and L.C.L. Hollenberg.
\newblock {A surface code quantum computer in silicon}.
\newblock {\em Science Advances}, 1(9):e1500707, 2015.

\bibitem{ONRMB16}
J.~O'Gorman, N.H. Nickerson, P.~Ross, J.J.L. Morton, and S.C. Benjamin.
\newblock {A silicon-based surface code quantum computer}.
\newblock {\em npj Quantum Information}, 2:16014, 2016.

\bibitem{B14}
R.~Barends, J.~Kelly, A.~Megrant, A.~Veitia, D.~Sank, E.~Jeffrey, T.C. White,
  J.~Mutus, A.G. Fowler, B.~Campbell, Y.~Chen, Z.~Chen, B.~Chiaro,
  A.~Dunsworth, C.~Neill, P.~O`Malley, P.~Roushan, A.~Vainsencher, J.~Wenner,
  A.N. Korotkov, A.N. Cleland, and J.M. Martinis.
\newblock {Logic gates at the surface code threshold: Superconducting qubits
  poised for fault-tolerant quantum computing}.
\newblock {\em Nature (London)}, 508:500--503, 2014.

\bibitem{L16}
C.J. Ballance, T.P. Harty, N.M. Linke, M.A. Sepiol, and D.M. Lucas.
\newblock {High-fidelity quantum logic gates using trapped-ion hyperfine
  qubits}.
\newblock {\em arxiv:1512.04600}, 2016.

\bibitem{PFW16}
A.~Paler, A.G. Fowler, and R.~Wille.
\newblock {Reliable quantum circuits have defects}.
\newblock {\em XRDS}, 23(1):34--38, 2016.

\bibitem{G98+}
D.~Gottesman.
\newblock {The Heisenberg Representation of Quantum computers}.
\newblock {\em quant-ph/9807006}, 1998.

\bibitem{DFTMN10}
S.J. Devitt, A.G. Fowler, T.~Tilma, W.J. Munro, and K.~Nemoto.
\newblock {Classical Processing Requirements for a Topological Quantum
  Computing System}.
\newblock {\em Int. J. Quant. Inf.}, 8:1--27, 2010.

\bibitem{GP10}
G.~Duclos-Cianci and D.~Poulin.
\newblock {Fast Decoders for Topological Quantum Codes}.
\newblock {\em Phys. Rev. Lett.}, 104:050504, 2010.

\bibitem{GP14}
G.~Duclos-Cianci and D.~Poulin.
\newblock {Fault-Tolerant Renormalization Group Decoded for Abelian Topological
  Codes}.
\newblock {\em Quant. Inf. Comp.}, 14:0721, 2014.

\bibitem{F15+}
A.G. Fowler.
\newblock {Minimum weight perfect matching of fault-tolerant topological
  quantum error correction in average $O(1)$ parallel time}.
\newblock {\em Quant. Inf. Comp.}, 15(0145-0158), 2015.

\bibitem{PF13}
A.G.~Fowler A.~Paetznick.
\newblock {Quantum circuit optimization by topological compaction in the
  surface code}.
\newblock {\em arXiv:1304.2807}, 2013.

\bibitem{DN12}
S.J. Devitt and K.~Nemoto.
\newblock Programming a topological quantum computer.
\newblock {\em Test Symposium (ATS), 2012 IEEE 21st Asian}, pages pp 55--60,
  2012.

\bibitem{DSMN13}
S.J. Devitt, A.M. Stephens, W.J. Munro, and K.~Nemoto.
\newblock Requirements for fault-tolerant factoring on an atom-optics quantum
  computer.
\newblock {\em Nature Communications}, 4:2524, 2013.

\bibitem{FD12}
A.G. Fowler and S.J. Devitt.
\newblock {A bridge to lower overhead quantum computation}.
\newblock {\em arXiv:1209.0510}, 2012.

\bibitem{FDJ13}
A.G. Fowler, S.J. Devitt, and C.~Jones.
\newblock {Surface code implementation of block code state distillation}.
\newblock {\em Sci. Rep.}, 3:1939, 2013.

\bibitem{G06}
S.J. Gay.
\newblock {Quantum Programming Languages}.
\newblock {\em Mathematical Structures in Computer Science}, 16(04):581--600,
  2006.

\bibitem{GLRSV13}
A.S. Green, P.L. Lumsdaine, N.J. Ross, P.~Selinger, and B.~Valiron.
\newblock {Quipper: A Scalable Quantum Programming Language}.
\newblock {\em ACM SIGPLAN Notices}, 48(6):333--342, 2013.

\bibitem{WS14}
D.~Wecker and K.M. Svore.
\newblock {LIQUi|>: A Software Design Architecture and Domain-Specific Language
  for Quantum Computing}.
\newblock {\em arXiv:1402.4467}, 2014.

\bibitem{MD16}
R.~Van Meter and S.J. Devitt.
\newblock {Local and Distributed Quantum Computation}.
\newblock {\em arxiv:1605.06951}, 2016.

\bibitem{Z98}
C.~Zalka.
\newblock {Fast Versions of Shor's quantum factoring algorithm}.
\newblock {\em quant-ph/9806084}, 1998.

\bibitem{VI05}
R.~Van Meter and K.M. Itoh.
\newblock {Fast Quatum Modular Exponentiation}.
\newblock {\em Phys. Rev. A.}, 71:052320, 2005.

\bibitem{WC00}
R.~Cleve and J.~Watrous.
\newblock {Fast Parallel circuits for the quantum fourier transform}.
\newblock {\em {Proc. 41st Annual IEEE Symposium on Foundations of Computer
  Science (FOCS 2000)}}, pages 526--536, 2000.

\bibitem{AMMR12}
M.~Amy, D.~Maslov, M.~Mosca, and M.~Roetteler.
\newblock {A meet-in-the-middle algorithm for fast synthesis of depth-optimal
  quantum circuits}.
\newblock {\em Computer-Aided Design of Integrated Circuits and Systems, IEEE
  Transactions on}, 32:818--830, 2013.

\bibitem{WBCB14}
Dave Wecker, Bela Bauer, Bryan~K. Clark, Matthew~B. Hastings, and Matthias
  Troyer.
\newblock Gate-count estimates for performing quantum chemistry on small
  quantum computers.
\newblock {\em Physical Review A}, 90(2):022305--, 08 2014.

\bibitem{GS12}
B.~Giles and P.~Selinger.
\newblock {Exact synthesis of multiqubit Clifford+T circuits}.
\newblock {\em Phys. Rev. A.}, 87:032332, 2013.

\bibitem{J13+}
N.~Cody Jones.
\newblock {\em {Logic Synthesis for Fault-Tolerant Quantum Computers}}.
\newblock PhD thesis, Ph.D Thesis, Stanford University, 2013.

\bibitem{J13+++}
Cody Jones.
\newblock Low-overhead constructions for the fault-tolerant toffoli gate.
\newblock {\em Phys. Rev. A}, 87:022328, Feb 2013.

\bibitem{RS14}
N.J. Ross and P.~Selinger.
\newblock {Optimal ancilla-free Clifford+T approximation of z-rotations}.
\newblock {\em arXiv:1403.2975}, 2014.

\bibitem{GKMR14}
D.~Gosset, V.~Kliuchnikov, M.~Mosca, and V.~Russo.
\newblock {An algorithm for the T-count}.
\newblock {\em Quant. Inf. Comp.}, 14:1277--1301, 2014.

\bibitem{KMM13}
V.~Kliuchnikov, D.~Maslov, and M.~Mosca.
\newblock {Asymptotically optimal approximation of single qubit unitaries by
  Clifford and T circuits using a constant number of ancillary qubits}.
\newblock {\em Phys. Rev. Lett.}, 110:190502, 2013.

\bibitem{PPND15}
A.~Paler, I.~Polian, K.~Nemoto, and S.J. Devitt.
\newblock {A Compiler for Fault-Tolerant High Level Quantum Circuits}.
\newblock {\em arxiv:1509.02004}, 2015.

\bibitem{PDF16}
A.~Paler, S.J. Devitt, and A.G. Fowler.
\newblock {Synthesis of Arbitrary Quantum Circuits to Topological Assembly}.
\newblock {\em arxiv:1604.08621}, 2016.

\bibitem{BH12}
Sergey Bravyi and Jeongwan Haah.
\newblock Magic-state distillation with low overhead.
\newblock {\em Phys. Rev. A}, 86:052329, Nov 2012.

\bibitem{J13++}
Cody Jones.
\newblock Multilevel distillation of magic states for quantum computing.
\newblock {\em Phys. Rev. A}, 87:042305, Apr 2013.

\bibitem{BK05+}
S.~Bravyi and A.~Kitaev.
\newblock {Universal Quantum Computation with ideal Clifford gates and noisy
  ancillas}.
\newblock {\em Phys. Rev. A.}, 71:022316, 2005.

\bibitem{Bin00}
The three-dimensional bin packing problem.
\newblock {\em Operations Research}, 48(2):256--267, 2000.

\bibitem{foldit}
Seth Cooper, Firas Khatib, Adrien Treuille, Janos Barbero, Jeehyung Lee,
  Michael Beenen, Andrew Leaver-Fay, David Baker, Zoran Popovic, and Foldit
  players.
\newblock Predicting protein structures with a multiplayer online game.
\newblock {\em Nature}, 466(7307):756--760, 08 2010.

\bibitem{eyewire}
Jinseop~S. Kim, Matthew~J. Greene, Aleksandar Zlateski, Kisuk Lee, Mark
  Richardson, Srinivas~C. Turaga, Michael Purcaro, Matthew Balkam, Amy
  Robinson, Bardia~F. Behabadi, Michael Campos, Winfried Denk, H.~Sebastian
  Seung, and the EyeWirers.
\newblock Space-time wiring specificity supports direction selectivity in the
  retina.
\newblock {\em Nature}, 509(7500):331--336, 05 2014.

\end{thebibliography}
\end{document}